\documentclass[a4paper,onecolumn,aps,showpacs,showkeys,nofootinbib,preprintnumbers,superscriptaddress,amsmath,amssymb,amsfonts]{revtex4}
\usepackage{graphicx}
\usepackage{amsmath}
\usepackage[latin1]{inputenc}
\usepackage[english]{babel}
\usepackage[T1]{fontenc}
\usepackage{amssymb}
\usepackage{amsfonts}
\usepackage{epsfig}
\usepackage{colordvi}
\usepackage{psfrag}
\usepackage{color}
\usepackage{dcolumn}
\usepackage{multirow}
\usepackage{hyperref}
\usepackage{epstopdf}
\usepackage{times}
\hypersetup{
  colorlinks=true,        % false: boxed links; true: colored links
  linkcolor=blue,         % color of internal links
  citecolor=magenta,      % color of links to bibliography
}

\begin{document}
\title{Investigating  dark energy equation of state with high redshift Hubble diagram}
%\title{Gauss-Bonnet cosmography}
\author{Marek Demianski}
\affiliation{Institute for Theoretical Physics, University of Warsaw, Pasteura 5, 02-093 Warsaw, Poland }
\affiliation{Department of Astronomy, Williams College, Williamstown, MA 01267, USA}
%\affiliation{Donostia International Physics Center (DIPC), 20018 Donostia-San Sebastian (Gipuzkoa) Spain}
\author{Elisabeta Lusso}
\affiliation{Dipartimento di Fisica e Astronomia, Universit\`a di Firenze, Via G. Sansone 1, 50019 Sesto Fiorentino, Firenze, Italy}
\affiliation{INAF-Osservatorio Astrofisico di Arcetri, 50125 Florence, Italy}
%\affiliation{Lab.Theor.Cosmology,Tomsk State University of Control Systems and Radioelectronics(TUSUR), 634050 Tomsk, Russia}
%\affiliation{$^{4}$Tomsk State Pedagogical University, Tomsk, ul. Kievskaya 60, 634061 Russian Federation}
\author{Maurizio Paolillo}
\affiliation{Dipartimento di Fisica, Universit\'a
di Napoli {}``Federico II'', Compl. Univ. di
Monte S. Angelo, Edificio G, Via Cinthia, I-80126, Napoli, Italy}
\affiliation{INFN Sezione  di Napoli, Compl. Univ. di
Monte S. Angelo, Edificio G, Via Cinthia, I-80126, Napoli, Italy}
\affiliation{INAF-Osservatorio Astronomico di Capodimonte, Via Moiariello 16, 80131 Napoli, Italy}
%\affiliation{Lab.Theor.Cosmology,Tomsk State University of Control Systems and Radioelectronics(TUSUR), 634050 Tomsk, Russia.}
\author{Ester Piedipalumbo}
\affiliation{Dipartimento di Fisica, Universit\'a
di Napoli {}``Federico II'', Compl. Univ. di
Monte S. Angelo, Edificio G, Via Cinthia, I-80126, Napoli, Italy}
\affiliation{INFN Sezione  di Napoli, Compl. Univ. di
Monte S. Angelo, Edificio G, Via Cinthia, I-80126, Napoli, Italy}
\author{Guido Risaliti}
\affiliation{Dipartimento di Fisica e Astronomia, Universit\`a di Firenze, Via G. Sansone 1, 50019 Sesto Fiorentino, Firenze, Italy}
\affiliation{INAF-Osservatorio Astrofisico di Arcetri, 50125 Florence, Italy}

\date{\today}

\begin{abstract}
Several  independent cosmological data, collected within the last twenty years,  revealed the accelerated expansion rate of the Universe, usually assumed to be
driven by the so called dark energy, which, according to recent estimates, provides now about $70\% $ of the total amount of matter-energy in the Universe. The nature of dark energy is yet unknown. Several models of dark energy have been proposed: a non zero cosmological constant, a potential energy of some self interacting scalar field, effects related to the non homogeneous distribution of matter, or effects due to alternative theories of gravity.
Recently,  it turned out that the standard flat $\Lambda$CDM is disfavored  (at $4\, \sigma$) when confronted with a high redshift Hubble diagram, consisting of supernovae of type Ia (SNIa), quasars  (QSOs) and gamma ray-bursts (GRBs) (\cite{Lusso16, Risaliti19, Lusso19}).
Here we use the same data to investigate if this tension is confirmed, using a different approach: actually in \citep{Lusso16, Risaliti19, Lusso19}, the deviation between the best fit model and the $\Lambda$CDM model was noticed by comparing cosmological parameters derived from cosmographic expansions of their theoretical predictions and observed high redshift Hubble diagram. In this paper we use a substantially  different approach, based on a specific parametrization of the redshift dependent equation of state (EOS) of dark energy component $w(z)$. Our statistical analysis is aimed to estimate the parameters characterizing the dark energy EOS: our results indicate (at $ >3\sigma$ level) an evolving dark energy EOS, while the cosmological constant $\Lambda$ has a constant EOS, $w_{\Lambda}=-1$. This result not only confirms the tension previously detected, but shows that it is not an artifact of cosmographic expansions.
 \end{abstract}
 \pacs{98.80.-k, 95.35.+d, 95.36.+x}
\keywords{Cosmology: observations, Quasars: general, Gamma-ray burst: general,Cosmology: dark energy, Cosmology: distance scale. }
 
\maketitle
%%%%%%%%%%%%%%%%%%%%%%%%%%%%%%%%%%%%%
\section{Introduction }\label{uno}
%%%%%%%%%%%%%%%%%%%%%%%%%%%%%%%%%%%%%%%

Recent observations of supernovae of type Ia (SNIa) indicate that the expansion rate of the Universe is accelerating  \citep{perl98, per+al99, Riess, Riess07,SNLS,Union2}. This unexpected result was confirmed by  analysis of small-scale anisotropies in temperature of the cosmic
microwave background radiation (CMBR) \citep{PlanckXIII}, and other cosmological data. The observed acceleration is due to so called
dark energy, that, in a fluid dynamics approach can be represented by a medium with a negative EOS, $\displaystyle{ w< - \frac{1}{3}}$. According to observational estimates, dark energy provides now about 70\% of matter energy in the Universe.\\However the nature of dark energy, is unknown. A large variety of models of dark energy have been introduced, including a cosmological constant \citep{carroll01}, or scalar field (see for instance \citep{SF,Peebles03}). \\The accelerated expansion of the Universe could be also expression of the inhomogeneous distribution of matter  (see for instance \citep{clark}),  or effects due to alternative theories of gravity.\\  Therefore the cause of the accelerated expansion of the Universe remains one of the most important open question in Cosmology and Physics: any new, independent measurement related to the expansion rate of the cosmological background, specially in different ranges of redshift, may shed new light on this topic and provide nontrivial test of the standard cosmological model (see \cite{Riess18a,Riess18b}). \\
Indeed, recently, other cosmological probes have entered the game: the first ones are  Ghirlanda et al. (2004), Fermiano et al.( 2005) using long GRBs ( \cite{Ghirlanda04}, and \cite{Firmani05}); Eisenstein et al. (2005) using the imprints of the BAOs in the large-scale structure (\cite{Eisenstein05} ); Chavez (2016) using HII galaxies (\cite{Chavez16}); Negrete et al. (2017) using extreme quasars \cite{Negrete17}.\\ Recently Risaliti \& Lusso \citep{Lusso16, Risaliti19} have shown that the combination of supernovae and quasars can extend and further constrain cosmological models. In
\citep{Lusso19}, some of us found  a tension between theoretical predictions of the $\Lambda$CDM model and a high-redshift Hubble diagram, on the basis of cosmographic expansions of the observable quantities.\\  Actually, cosmography allows to investigate the {\it kinematic} features of the evolution of the universe, assuming only that the space time geometry is described by the Friedman-Lemaitre-Robertson-Walker metric, and adopting  Taylor expansions, in the traditional approach, or  logarithmic polynomial expansions, in a generalized approach \citep{Lusso16, Risaliti19} , of basic cosmological observables. \\
From  observational data it is possible to constrain the cosmographic parameters, and their probability distributions. These constraints can provide information about the nature of dark energy  \citep{MGRB2}, and  are model independent, but depend on the properties of convergence of the cosmographic series. \\Here we approach this tension from a different point of view, and try to figure out from the observations whether the dark energy equation of state (EOS) evolve or not.  Actually it turns out that the  cosmological constant $\Lambda$ has a constant EOS, $w_{\Lambda}=-1$, whereas in most cosmological models, following a fluid dynamics approach, we can introduce at least an effective EOS of dark energy, depending on the redshift $z$.\\ This happens, for instance, in extended theories of gravity \citep{scalar-tensor1,scalar-tensor2,ft,GL}, or in an interacting quintessence cosmology \citep{Noetherint}: actually in the generalized Friedman equations, that drive the dynamics of these models, we can identify  {\it effective} density and pressure terms, and define {\it effective} EOS. Understanding from the data whether or not the dark energy EOS  is constant,  independently of any assumptions on its nature, is, therefore, a daunting yet rewarding task of probing the evolution of dark energy \citep{Hzinvestigation}. \\
Here we parameterize $w(z)$  using the Chevallier-Polarski-Linder (CPL) model \citep{cpl1,cpl2}. Noteworthy in this regard is that we do not  intend to use the CPL model to test its robustness in providing reliable reconstruction of dark energy models, as for instance in \cite{Linden05,Scherrer15}, but, rather, infer from high redshift data if the dark energy EOS is constant or not, and, therefore, if the expansion rate is compatible with the flat $\Lambda$CDM model. \\
Actually, despite the  $\Lambda$CDM enormous success, some tensions and problem have been detected:  a combined analysis of the Planck angular power spectra with different luminosity distance measurements are in strong disagreement with the flat $\Lambda$CDM (\cite{Di Valentino}). Moreover, it turns out that there are some tensions among the values of cosmological parameters inferred from independent datasets. The most famous and persisting one is related to the value of the Hubble constant $H_0 $ as measured by Planck and recently by \cite{Choi} with respect the value extracted from Cepheid-calibrated local distance ladder measurements (see for instance  \cite{Riess19}) ( the so called $H_0$ tension).\\
 Several papers present interesting attempt to solve this tension. For instance, Poulin et al. (\cite{Poulin19}) propose an early dark energy model to resolve the Hubble tension. They assume existence of a scalar field that adds dark energy equal to about $10\%$ of the energy density at the end of the radiation-dominated era at $z\simeq 3500$, and then it dilutes; after that the energy density components are the same as in the $\Lambda$CDM model. Another example is the consequence of string theory that predicts existence of an {\it axiverse}, i.e. a huge number of extremely light particles with very peculiar physical properties.  It seems that a simple modification of the physical properties of these particles is enough to accommodate the Hubble tension (\cite{Kamio14}). \\In our analysis we use the (SNIa) Hubble diagram (Union2.1 compilation), the gamma-ray burst (GRBs) and
QSOs Hubble diagram. We also use Gaussian priors on the distance from the Baryon Acoustic Oscillations (BAO), and the Hubble constant $h$, these priors have been included in order to help break the degeneracies among model parameters.  To constrain cosmological parameters we perform Monte Carlo Markov Chain (MCMC) simulations.
The structure of the paper is as follows. In Sect. 2 we describe the CPL parametrization used in our
analysis. In Sect. 3 we describe the data sets, and in Sect. 4 we describe the statistical
analysis and present our results.  General discussion of our results and conclusions are presented in Sect. 5.

\section{Parametrization of the dark energy EOS}

Available now observational data imply that the expansion rate of the Universe is accelerating. This accelerated expansion
is conveniently described by the standard $\Lambda$CDM cosmological model. However to test if the accelerated expansion is
due to the non zero cosmological constant it is necessary to consider a more general model of evolving dark energy that can be described by a simple equation of state (EOS)

\begin{eqnarray}
p_{de}(z) =w(z) \rho_{de}(z)\,, \label{eq: eos}
\end{eqnarray}

where $\rho_{de}$ is the effective energy density of dark energy and $p_{de}(z)$ is its pressure. The proportionality
coefficient $w(z)$ determines the dark energy EOS. When $w(z) = - 1$ the cosmological constant plays the role of dark energy.

In the standard spatially flat Friedman-Lemaitre-Robertson-Walker model the scale factor $a(t)$ is determined by the Friedman equations:

\begin{equation}
H^2 = (\frac{\dot a}{ a})^{2} =  \frac{8 \pi G}{3} (\rho_m + \rho_{de}) \,, \label{fried2}
\end{equation}

\begin{equation}
\frac{\ddot{a}}{a} = - \frac{4 \pi G}{3} \ \left(\rho_m + \rho_{de} + 3 p_{de}\right) \,, \label{fried1}
\end{equation}

where $H$ is the Hubble parameter, the dot denotes the derivative with respect to the cosmic time $t$, and $\rho_{m}$ is the density of non relativistic matter. When the Universe is filled in with other non interacting matter components
their energy density $\rho_{i}$ and pressure $p_{i}$ are related by EOS for each component of the cosmological fluid
$p_{i} = w_{i}\rho_{i} $ and $i = 1, ..., n$. Non relativistic matter is usually considered to be pressure less so it
is characterized by $w = 0$. The cosmological constant can be treated as a medium with $w = - 1$. The non interacting matter components satisfy the continuity equation

\begin{equation}
\frac{\dot{\rho_i}}{\rho_i} = - 3 H \left(1 +
\frac{p_i}{\rho_i}\right) = - 3 H  \left[1 + w_i(t)\right]\,.
\label{eq: continuity}
\end{equation}

In the later stages of evolution of the Universe its dynamics is determined only by two components:
the non relativistic matter and dark energy. In this case the Friedman equation (\ref{fried2}) can be rewritten as
\begin{eqnarray}
  H(z,{\mathrm \theta}) &=& H_0 \sqrt{\Omega_m (z+1)^3 + (1-\Omega_m) g(z, {\mathrm \theta})}\,,
\label{heos}
\end{eqnarray}

where $\Omega_{m} = \frac{8 \pi G \rho_{m}}{3 H_{0}^{2}}$ is the matter density parameter,
 and
$g(z, {\mathrm \theta}) = \frac{
\rho_{de}(z)}{\rho_{de}(0)} = \exp{(3 \int_0^z \frac{w(x,{\mathrm \theta})+1}{x+1} \, dx)}$. Here $w(z,{\mathrm
\theta})$ characterizes the dark energy EOS and ${\mathrm \theta} = ({\mathrm \theta}_{1}, ..., {\mathrm
\theta}_{n})$ are additional parameters that describe different models of dark energy. Using the Hubble parameter
(\ref{heos}) we define the luminosity distance in the standard way as

\begin{eqnarray}
% \nonumber to remove numbering (before each equation)
d_L(z,{\mathrm \theta}) &=&\frac{c}{H_0} (1+z)\int^{z}_{0}\frac{d\zeta}{ H(\zeta, {\mathrm \theta} )}\, .
\label{lumdgen}
\end{eqnarray}

 Moreover we can define the angular diameter distance and the volume distance as:
\begin{eqnarray}
  d_A(z,{\mathrm \theta})  &=&  \frac{c}{H_0}\frac{1}{1+z} \int_0^z \frac{d\zeta}{ H(\zeta, {\mathrm \theta} )}\,,\label{angd} \\
  d_V(z,{\mathrm \theta})  &=& \left[\left(1+z\right) d_A(z,{\mathrm \theta})^2\frac{c z}{H(z,{\mathrm \theta})}\right]^{\frac{1}{3}}\,.\label{volumed}
\end{eqnarray}

To compare predictions of different models of dark energy with high redshift Hubble diagram we use the standard definition of the distance modulus:

\begin{equation}
\mu(z, {\mathrm \theta}) = 25 + 5 \log{d_L(z,{\mathrm \theta})}\,.
\label{eq:defmu}
\end{equation}

Several different models of dark energy were proposed so far. Here to simplify our analysis we use
a two parametric EOS proposed by Chevalier, Polarski and Linder (CPL)

\begin{equation}
w(z) =w_{0} + w_{1} \frac{z}{ 1 + z} \,,
\label{cpleos}
\end{equation}

where $w_{0}$ and $w_{1}$ are constants to be determined by the fitting procedure. From the assumed form of
$w(z)$ it is clear that

\begin{eqnarray}
&&\lim_{z \to 0}w(z)=w_0\,,\\
&&\lim_{z \to \infty}w(z)=w_0+w_{1}\,.
\end{eqnarray}

To check if properties of dark energy change in time it is necessary to use high redshift data to find out
if $w_{1}$ is zero or not.

\begin{figure}
\begin{center}
\includegraphics[width=8cm]{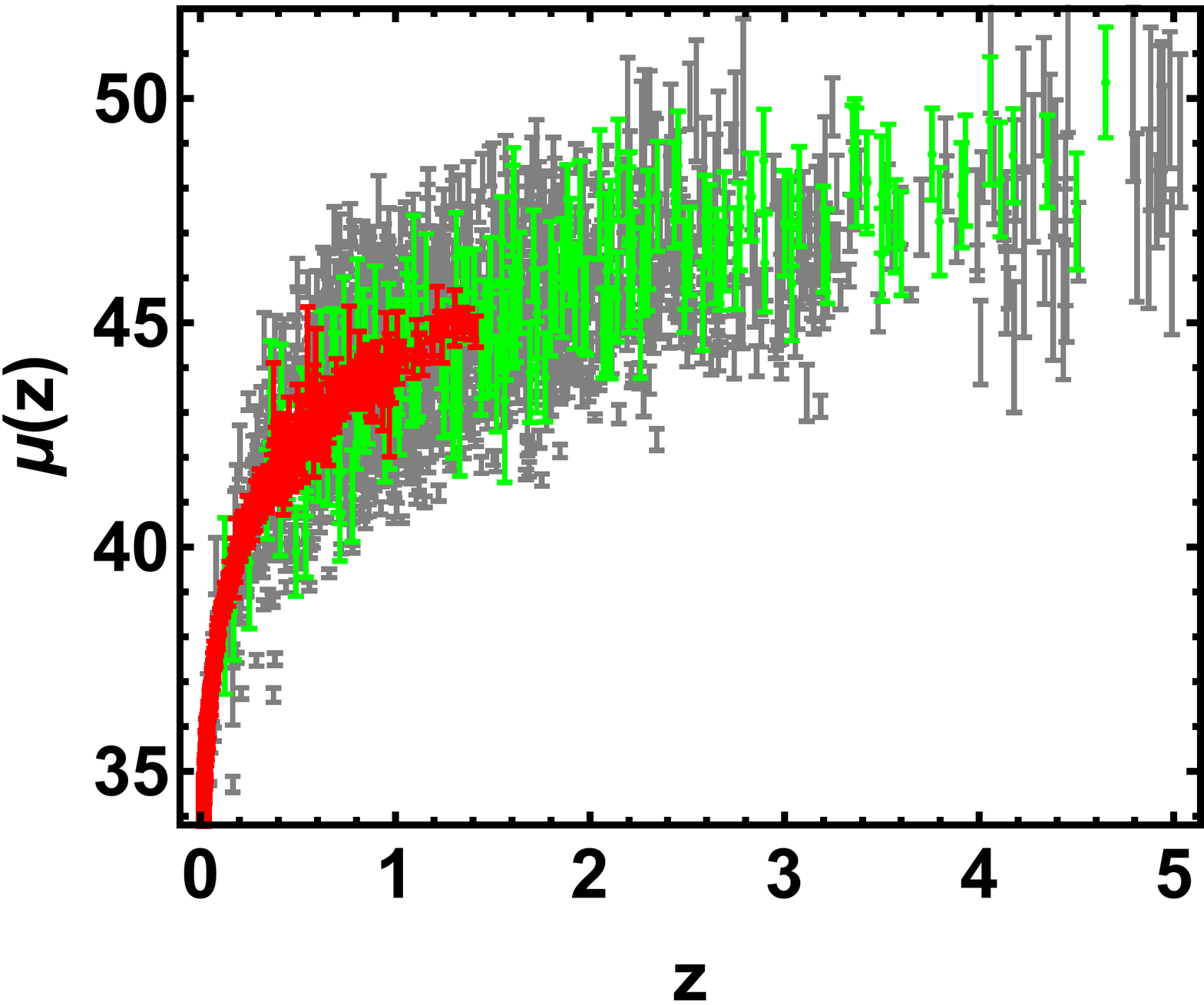}% This is a *.eps file
\end{center}
\caption{ Hubble diagram of SNIa (red points), quasars (gray points) and GRBs (green points), with their respective $1 \sigma$ uncertainties.
}\label{data}
\end{figure}

\section{The data sets}

To perform our analysis and determine the basic cosmological parameters, we construct a Hubble diagram by combining four data samples. In our analysis we build up a Hubble diagram spanning a wide redshift range. To this aim we combine different data sets.
\\

\subsection{Supernovae Ia}

 SNIa observations gave the first strong indication of the recent accelerating
expansion of the Universe (  \cite{per+al99} and \cite{Riess}). In our analysis we use the
 recently updated Supernovae Cosmology Project Union 2.1
compilation \cite{Union2.1}, containing $580$ SNIa, spanning the redshift
range ($0.015 \le z \le 1.4$). We compare the theoretical
 distance modulus $\mu(z)$, based on the definition of the
distance modulus in different cosmological models:

\begin{equation}
\mu(z_{j}) = 5 \log_{10} ( D_{L}(z_{j}, \{\theta_{i}\}) )+\mu_0\,,
\end{equation}
where $D_{L}(z_{j}, \{\theta_{i}\})$ is the Hubble free
luminosity distance, and $\theta_{i}$ indicates the set of
parameters that appear in different dark energy equations of
state considered in our analysis. The parameter $\mu_{0}$
encodes the Hubble constant and the absolute magnitude $M$.
We used an alternative version of the $\chi^2$:
\begin{equation}
\label{eq: sn_chi_mod}
\tilde{\chi}^{2}_{\mathrm{SN}}(\{\theta_{i}\}) = c_{1} -
\frac{c^{2}_{2}}{c_{3}}\,,
\end{equation}
where
\begin{equation}
c_{1} = \sum^{ N_{SNIa}}_{j = 1} \frac{\left(\mu(z_{j}; \mu_{0}=0,
\{\theta_{i}\} )- \mu_{obs}(z_{j})\right)^{2}}{\sigma^{2}_{\mathbf{\mu},j}}\, ,
\end{equation}

\begin{equation*}
c_{2} = \sum^{N_{SNIa}}_{j = 1} \frac{\left(\mu(z_{j}; \mu_{0}=0,\{\theta_{i}\} -
\mu_{obs}(z_{j})\right)^2}{\sigma^{2}_{\mathbf{\mu},j}}\, ,
\end{equation*}

\begin{equation}
c_{3} = \sum^{N_{SNIa}}_{j = 1}
\frac{1}{\sigma^{2}_{\mathbf{\mu},j}}\,.
\end{equation}
It is worth noting that
\begin{equation}
\chi^{2}_{\mathrm{SN}}(\mu_{0}, \{\theta_{i}\}) = c_{1} - 2 c_{2}
\mu_{0} + c_{3} \mu^{2}_{0} \,,
\end{equation}
which clearly has a minimum for $\mu_{0} = \displaystyle\frac{c_{2}}{c_{3}}$, so
 that
 \begin{equation}
\tilde{\chi}^{2}_{\mathrm{SN}} \equiv
\chi^{2}_{\mathrm{SN}}(\mu_{0} =\frac{c_{2}}{c_{3}}, \{\theta_{i}\})\,.\nonumber
\label{chisquared}
\end{equation}
\subsection{Gamma-ray burst Hubble diagram}
Gamma-ray bursts are visible up to high redshifts thanks to the enormous released energy, and therefore are  good candidates  for our high-redshift cosmological investigations.
They show non thermal spectra which can be empirically modeled with the  Band function, i.e., a smoothly broken power law with parameters: the low-energy spectral index $\alpha$, the high energy spectral index $\beta$ and the {\it roll-over} energy $E_0$. Their spectra show a peak corresponding to a specific (and observable) value of the photon energy $E_{\rm p} = E_0 (2 + \alpha)$; indeed it turns out that for GRBs with measured spectrum and redshift it is possible to evaluate the intrinsic peak energy, $E_{\rm p,i} = E_{\rm p} (1 + z)$ and the isotropic equivalent radiated energy
\begin{equation}
E_{\rm iso}= 4 \pi d_L(z,{\mathrm \theta}) \left(1+z\right)^{-1}\int^{10^4/(1+z)}_{1/(1+z)} E N(E)
dE\,,
\label{eqEiso}
\end{equation}

where $N(E)$ is the Band function:
\[N(E)=\left\{
\begin{array}{ll}
 A \left(\frac{E}{100keV}\right)^{\alpha} \exp{\left(\frac{-E}{E_0}\right)}\,,&\left(\alpha-\beta\right)\geq 0\,,\\
 A \left(\frac{\left(\alpha-\beta\right)E}{100keV}\right)^{\alpha-\beta} \exp{\left(\alpha-\beta\right)\left(\frac{E}{100keV}\right)^{\beta}}\,,&\left(\alpha-\beta\right)E_0\leq E\,,\\
\end{array}
\right. \]
 $E_{\rm p,i} $ and $E_{\rm iso} $ span several orders of magnitude (GRBs cannot be considered standard candles), and show distributions approximated by Gaussians plus a tail at low energies.
However in 2002, based on a small sample of BeppoSAX, it turned out that $E_{\rm p,i} $ is strongly correlated with $E_{\rm iso}$ \cite{Amati02}.
This correlation, commonly called Amati relation within the GRB community, has been confirmed in subsequent observations and provide a reliable instrument  to standardize GRBs as a distance indicator, in a way similar to t the Phillips relation to standardize SNIa ( see, for instance, \cite{Amati08},\cite{MGRB1,MGRB2}  and reference therein).
It is clear from Eq. (\ref{eqEiso}) that, in order to get the isotropic equivalent radiated energy, $E_{iso} $ it is necessary to specify the fiducial cosmological model and its basic parameters. But we want to use the observed properties of GRBs
to derive the cosmological parameters. Several procedures to overcome this circular situation have been proposed (see for instance \cite{S07,BP08,Amati19,MEC11}). 

Here we performed our analysis using a GRB Hubble diagram data set obtained by  calibrating the $E_{\rm p,i}$-- $E_{\rm iso}$ correlation on a SNIa data (\cite{MGRB2,MGRB1}). Actually we applied a local regression technique to estimate the distance modulus $\mu(z)$ from
the  SCP Union2 compilation. To obtain an estimation of $\mu(z_i)$ we  order the SNIa dataset  according to increasing values of $|z - z_i|$. Therefore we select the first $n = \alpha N_{SNIa}$, where $\alpha$  is a user selected value and $N_{SNIa}$ the total number of SNIa. Then we fit a first order polynomial to these data, weighting each SNIa with the corresponding value of an appropriate weight function, like, for instance
\begin{equation}
W(u) = \left \{
\begin{array}{ll}
(1 - |u|^2)^2 & |u| \le 1 \\ ~ & ~ \\ 0 & |u| \ge 1\,.
\end{array}
\right .
\label{eqWeight}
\end{equation}
The zeroth order term is the best estimate of $\mu(z)$. The error on $\mu(z)$ is provided by the root mean square of the weighted
residuals with respect to the best fit value. In Eq. (\ref{eqWeight})  $u = |z - z_i|/\Delta$ and $\Delta$ indicates the maximum value of the $|z - z_i|$ over the chosen subset. Having estimated the distance modulus
at redshift $z$ in a model independent way, we can fit
the $E_{\rm p,i}$ -- $E_{\rm iso}$  correlation  using the \textit{local regression reconstructed}
$\mu(z)$ in Eq. (\ref{eqEiso}). It is worth noting that we already discussed some aspects of this topic in our previous papers \cite{MEC11},\cite{MGRB1}, where, apart from other issues,  we described how  it is
possible to simultaneously constrain the calibration parameters
and the set of cosmological parameters: we found that  the calibration parameters are fully consistent with the
results obtained from the SNIa calibrated data.

In order to investigate a possible $z$ dependence
of the correlation coefficients, in the calibration procedure we added terms representing the $z$-evolution, which
are  assumed to be  power-law functions: $g_{iso}(z)=\left(1+z\right)^{k_{iso}}$ and
$g_{p}(z)=\left(1+z\right)^{k_{p}}$ (\citep[][]{MGRB1}). Therefore  $E_{\rm iso}^{'}
=\displaystyle\frac{E_{\rm iso}}{g_{iso}(z)}$ and $E_{\rm p,i}^{'} =\displaystyle\frac{E_{\rm p,i}}{g_{p}(z)}$ are the
de-evolved  quantities included in a 3D correlation:
\begin{eqnarray}
 \label{eqamatievol}
&&\log \left[\frac{E_{\rm iso}}{1\;\mathrm{erg}}\right] = b+a \log  \left[
    \frac{E_{\mathrm{p,i}} }{300\;\mathrm{keV}} \right]+\nonumber \\&& + \left(k_{iso} - a \,k_{p}\right)\log\left(1+z\right)\,.
\end{eqnarray}
This de-evolved correlation was calibrated applying the same  local regression
technique  previously adopted (\cite{MGRB1,MGRB2} ), but considering a 3D Reichart
likelihood:
\begin{eqnarray}
 \label{reich3dl}
&&L^{3D}_{Reichart}(a, k_{iso}, k_{p}, b,  \sigma_{int})=\nonumber\\&& \frac{1}{2} \frac{\sum{\log{(\sigma_{int}^2 + \sigma_{y_i}^2 + a^2
\sigma_{x_i}^2)}}}{\log{(1+a^2)}}\nonumber \\ &+& \frac{1}{2} \sum{\frac{(y_i - a x_i -(k_{iso}-\beta) z_i-b)^2}{\sigma_{int}^2 + \sigma_{x_i}^2 + a^2
\sigma_{x_i}^2}}\,,
\end{eqnarray}
where $\beta= a\, k_{p}$. We also used the MCMC method to maximize the likelihood and ran
five parallel chains and the Gelman-Rubin convergence test. We
found that $a=1.87^{+0.08}_{-0.09}$, $k_{iso}=-0.04\pm 0.1$;
$\beta=0.02\pm 0.2$\,; $\sigma_{int}=0.35_{-0.03}^{+0.02}$, so
that $b= 52.8_{-0.06}^{+0.03}$. After fitting the  correlation
and estimating its parameters, we used them to construct the GRB
Hubble diagram. Detailed discussion of the GRBs sample and possible selection effects is presented also in (\cite{Amati08, A-DV,MGRB1}).
\\

\subsection{Quasars}
A physical relation has been observed between the optical-UV disk and the X-ray corona consisting in  a log-log relation between their respective fluxes. From previous studies, Lusso et al.  (see \cite{Lusso16}) found out  a dispersion varying between 0.35 to 0.4 dex in this correlation. Their first sample was further reduced by eliminating quasars with host galaxy contamination, reddening, X-ray obscured objects and radio loudness  (Lusso \& Risaliti, 2016) to reach a dispersion of 0.21-0.24 dex.
The quasar sample used here is presented in \citep{Risaliti19, Lusso19} and consists of 1598 sources in the redshift range $0.04 < z < 5.1$ . Distance moduli have been estimated by calibrating the power-law correlation between the ultraviolet and the X-ray emission observed in quasars \citep{Lusso16,Risaliti19, Lusso19}).
More details on the sample are provided in \cite{Lusso19}.
  \\
\subsection{H(z) data}
The Hubble parameter depends on the differential age of the Universe and can be measured  using the cosmic chronometers: usually, $dz$ is obtained from spectroscopic surveys, and, if cosmic chronometers are identified, we can measure $dt$, in the  redshift interval $dz$. We used a list of $28$ $H(z)$ measurements, compiled in \cite{farooqb}. The Hubble parameter H(z) can be measured  through the differential age technique, based on passively evolving red galaxies as cosmic chronometers. Actually it turns out that  H(z) depends on the differential variation of the cosmic time with redshift according to the following relation:
\begin{equation}
H(z)=- \frac{1}{1+z} \left( \frac{dt}{dz}\right)^{-1}\,.
\label{hz}
\end{equation}
The  term $\displaystyle \left( \frac{dt}{dz}\right)$, is estimated
from the age of old stellar populations in red galaxies from their high resolution spectra.
\\
\subsection{BAOs}

In order to reduce the degeneracy among cosmological parameters we also use some constraints from standard rulers. Actually the BAOs, which are related to imprints of the primordial acoustic waves on the galaxy power spectrum, are widely used as such rulers.
In order to use BAOs as constraints, we follow \cite{P10} by first defining\,:
\begin{equation}
d_z = \frac{r_s(z_d)}{d_V(z)}\,,
\label{eq: defdz}
\end{equation}
with $z_d$ being the drag redshift, $d_V(z)$ the volume distance, and $r_s(z)$ the comoving sound horizon given by\,:
\begin{equation}
r_s(z) = \frac{c}{\sqrt{3}} \int_{0}^{(1 + z)^{-1}}{\frac{da}{a^2 H(a) \sqrt{1 + (3/4) \Omega_b/\Omega_{\gamma}}}} \,,
\label{defsoundhor}
\end{equation}
here $\Omega_{\gamma}$  is the radiation density parameter.  We fix  $r_s(z_d) = 152.6 \ {\rm Mpc}$  and the volume distance is defined in Eq. (\ref{volumed}).
The values of $d_z$ at $z = 0.20$ and $z = 0.35$ have been estimated by Percival et al. (2010) using the SDSS DR7 galaxy sample (\cite{P10}) so that we define $\chi^2_{BAO} = {\bf D}^T {\bf C}_{BAO}^{-1} {\bf C}$ with ${\bf D}^T = (d_{0.2}^{obs} - d_{0.2}^{th}, d_{0.35}^{obs} - d_{0.35}^{th})$ and ${\bf C}_{BAO}$ is the BAO covariance matrix.

\section{Statistical analysis}
To compare the high-redshift data described above with the CPL parametrization, we use a Bayesian approach, and we
apply the  MCMC method to maximize the likelihood function ${\mathcal{L}}(\mathbf{p})$:
\begin{eqnarray}
&&{\mathcal{L}}(\mathbf{p})  \propto  \frac{\exp(-\chi^2_{GRB}/2)}{(2 \pi)^{\frac{{\mathcal{N}}_{GRB}}{2}} |{\bf
C}_{GRB}|^{1/2}} \frac{\exp(-(\chi^2_{QSO}/2)}{(2\pi)^{\frac{{\mathcal{N}}_{QSO}}{2}} |{\bf C}_{QSO}|^{1/2}}\nonumber\\
&&\times   \frac{\exp(-\chi^2_{SNIa}/2)}{(2 \pi)^{\frac{{\mathcal{N}}_{SNIa}}{2}} |{\bf C}_{SNIa}|^{1/2}}
\frac{\exp(-\chi^2_{BAO}/2)}{(2 \pi)^{{\mathcal{N}}_{BAO}/2} |{\bf C}_{BAO}|^{1/2}}\nonumber\\ &&\times
\frac{\exp(-\chi^2_{H}/2)}{(2 \pi)^{{\mathcal{N}}_{H}/2} |{\bf C}_{H}|^{1/2}}. \, \label{defchiall}
\end{eqnarray}

In the Eq. (\ref{defchiall})

\begin{equation}
\chi^2(\mathbf{p}) = \sum_{i,j=1}^{N} \left( x_i - x^{th}_i(\bf p)\right)C^{-1}_{ij}  \left( x_j - x^{th}_j(\bf
p)\right) \,, \label{eq:chisq}
\end{equation}
where $\mathbf{p}$ denotes the parameters that determine the cosmological model, $N$ is the number of
data point. $C_{ij}$ is the covariance matrix ( ${\mathbf C}_{SNIa/GRB/QSO/H}$ indicates the SNIa/GRBs/QSO/H  covariance
matrix). The term  $\displaystyle \frac{\exp(-\chi^2_{H}/2)}{(2 \pi)^{{\mathcal{N}}_{H}/2} |{\bf C}_{H}|^{1/2}}  $ in
Eq. (\ref{defchiall}) is the likelihood relative to $H(z)$ . It is worth noting that by joining the GRB and QSO Hubble
diagrams we can probe the background cosmological expansion over a redshift range more appropriate for
studying dark energy than the one covered by SNIa only.
We used a Metropolis-Hastings algorithm: we start from an initial parameter vector ${\mathbf \theta^{start}} $, and we generate a new trial vector $ {\mathbf \theta^{next}}$ from a tested density $f({\mathbf {\theta^{next}}, {\theta^{start}}})$, which represents the conditional
probability of $ {\mathbf \theta^{next}}$ , given  $ {\mathbf {\theta^{start}}}$.  The probability of accepting the new vector  $ {\mathbf {\theta^{next}}}$ is described by
\begin{equation}
F({\mathbf {\theta^{next}, \theta^{start}}}) = min \left\{1,
\frac{\mathcal{L}({\mathbf{d}}|{{\mathbf \theta^{next}}}) Pr({\mathbf{\theta^{next}}})
f({\mathbf {\theta^{next}}{\mathbf{\theta^{start}}}})}{\mathcal{L}({\mathbf{d}}|{{\mathbf \theta^{next}}}) Pr({\mathbf{\theta^{start}}})
f({\mathbf{ \theta^{next}}},{\mathbf{\theta^{start}}})}\right\}
\end{equation}
where ${\mathbf{d}}$ are the data, $\mathcal{L}({\mathbf{d}}|{\mathbf{p'}})
\propto \exp(-\chi^{2}/2)$ is the likelihood function,
$Pr({\mathbf{\theta}})$ is the prior on the parameters. Moreover we assume that  $f({\mathbf{\theta^{start}}}, {\mathbf{\theta^{next}}}) \propto \exp(-\Delta \theta^{2} / 2
\delta^{2})$, with $\Delta {\mathbf{\theta}} = \left({\mathbf{\theta^{start}}} -
{\mathbf{\theta^{next}}}\right)$, and the dispersion $\delta=15\%$ for any step.

\begin{figure}[h!]
\begin{center}
\includegraphics[width=8cm]{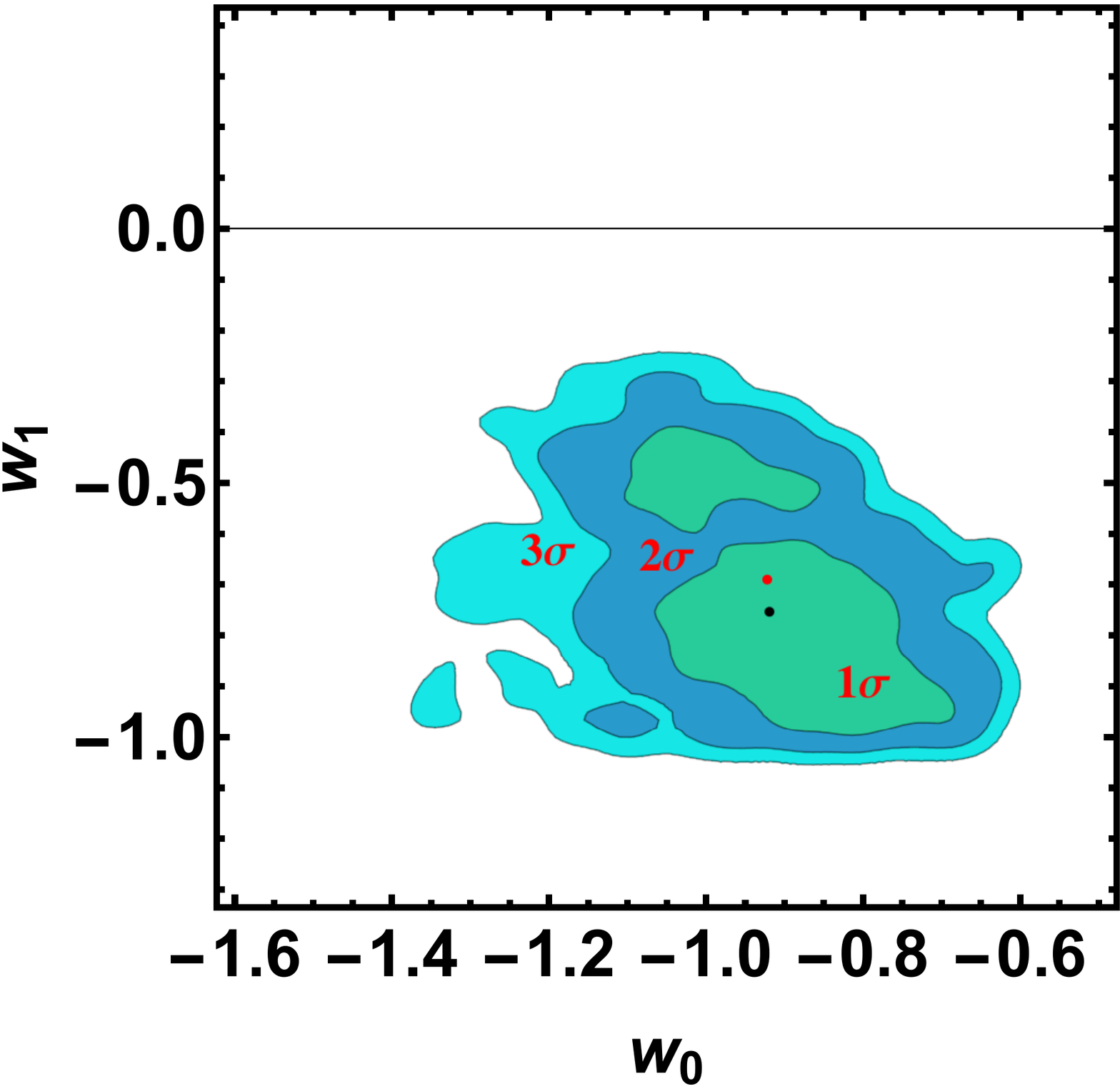}% This is a *.eps file
\end{center}
\caption{ 2D confidence regions in the $w_0-w_1$ plane for the CPL model, obtained from all the data sets. It turns out
that the case $w_0=-1\,, w_1=0$, corresponding to the standard $\Lambda$CDM model, is disfavored at $3 \sigma$, in full
agreement with the results of \citep{Lusso19} that use a different approach. The red and black points correspond to the best fit value and the mean respectively. }\label{fig1}
\end{figure}

\begin{figure}[h!]
\begin{center}
\includegraphics[width=8cm]{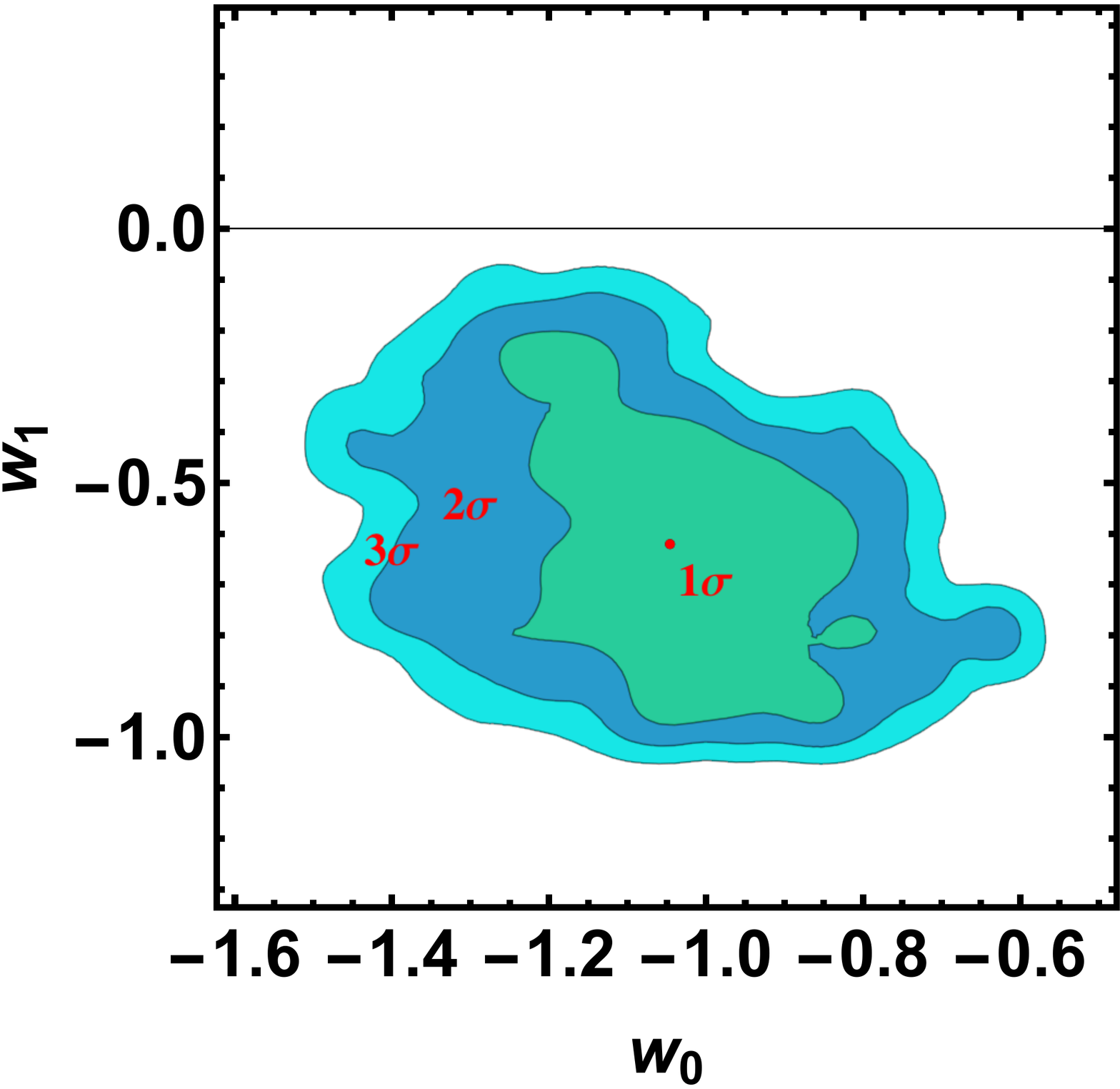}
\end{center}
\caption{2D confidence regions in the  $w_0-w_1$ plane for the CPL model, obtained from SNIa, GRB Hubble diagram and
the $H(z)$ measurements. It turns out that the case $w_0=-1\,, w_1=0$, corresponding to the $\Lambda$CDM model, is disfavored
at more than  $3 \sigma$. The red point corresponds to the best fit value and the mean respectively, that in this case are indistinguishable.}\label{fig2}
\end{figure}

\begin{figure}[h!]
\begin{center}
\includegraphics[width=8cm]{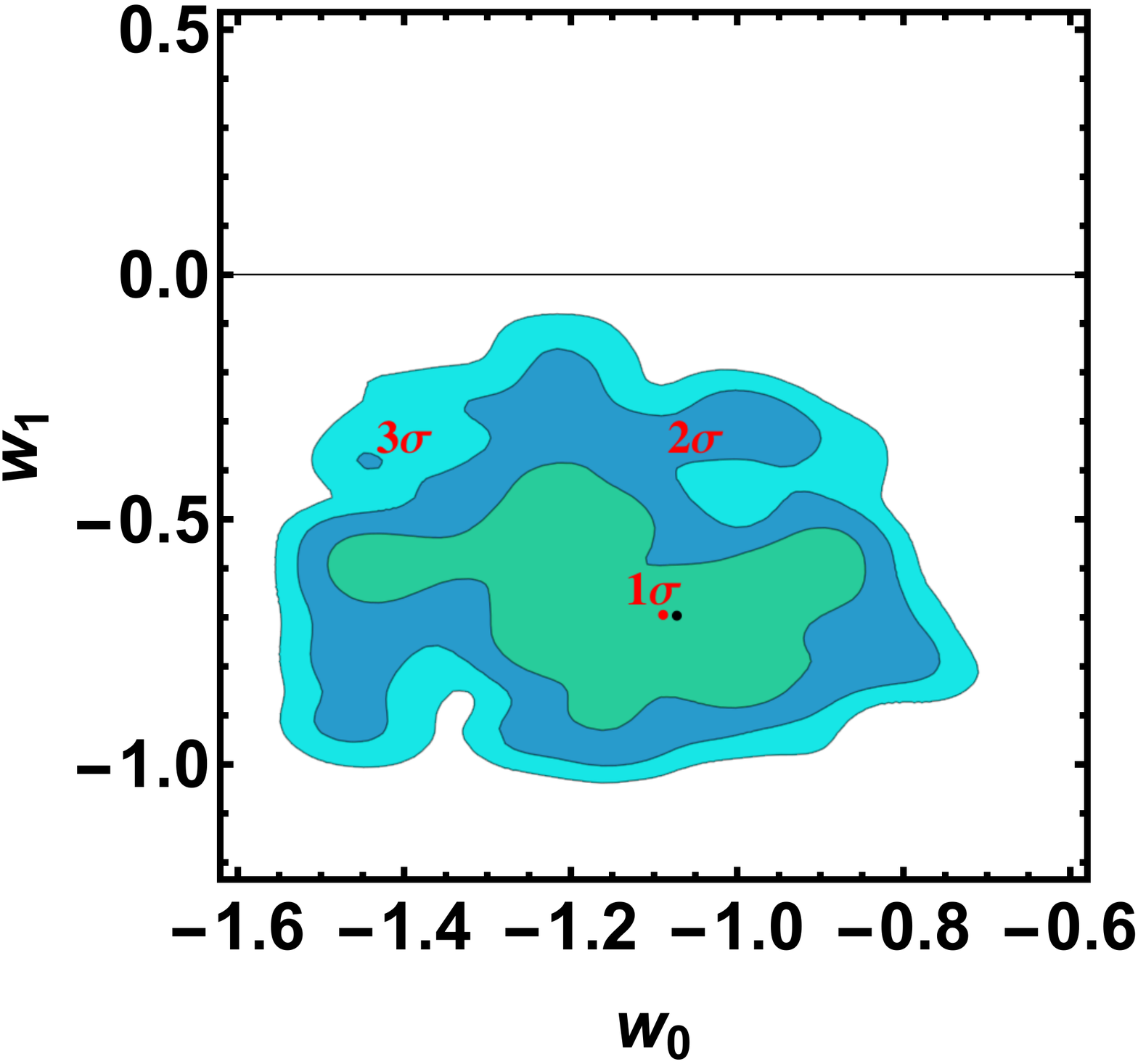}% This is a *.eps file
\end{center}
\caption{ 2D confidence regions in the $w_0-w_1$ plane for the CPL model, obtained from the QSO and GBR Hubble diagram: it turns out that the case $w_0=-1\,, w_1=0$, corresponding to the standard $\Lambda$CDM model, is disfavored at $3 \sigma$. The red and black points correspond to the best fit value and the mean respectively. }\label{fig3}
\end{figure}

\begin{figure}[h!]
\begin{center}
\includegraphics[width=9cm]{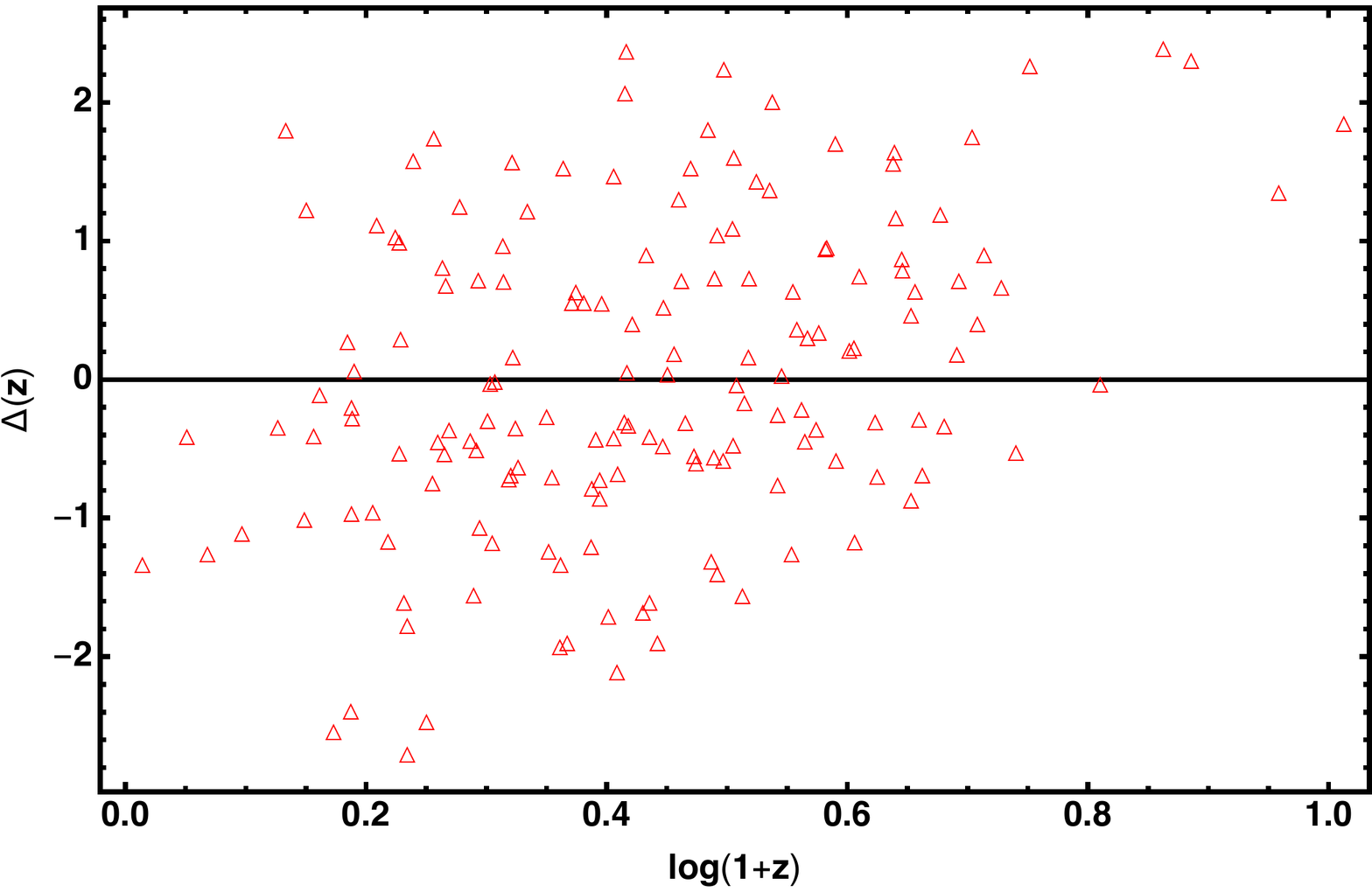}
\end{center}
\caption{Distribution of the Hubble diagram residuals as a function of z for the GRBs sample.}\label{residual}
\end{figure}

\begin{figure}[ht]
\begin{center}
\includegraphics[width=9cm]{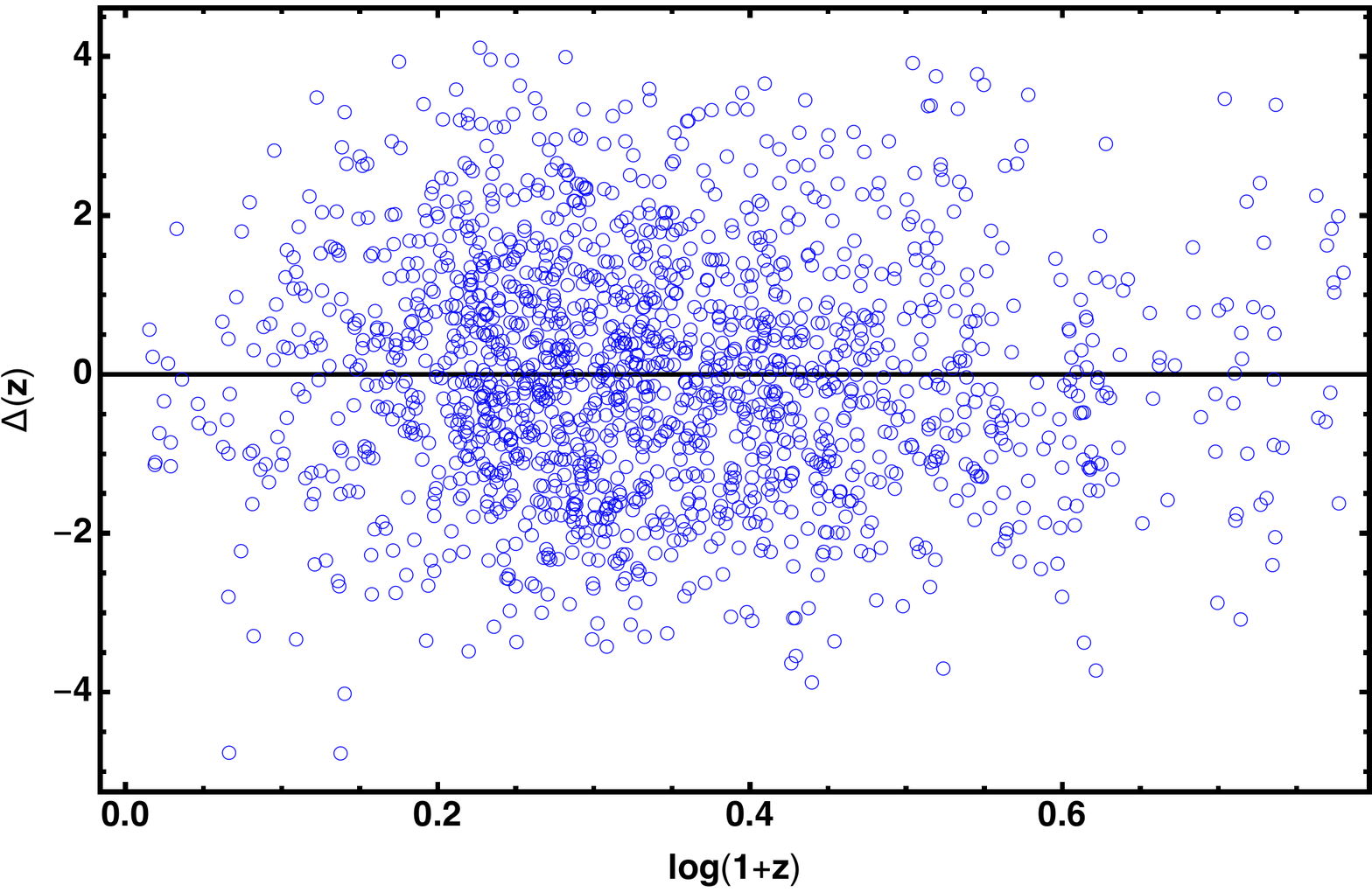}
\end{center}
\caption{Distribution of the Hubble diagram residuals as a function of z for the QSOs sample.}\label{residual2}
\end{figure}

In order to sample the space of parameters, we run three
parallel chains and use the Gelman-Rubin test to control the convergence of the chains. We use uniform priors for the parameters. After that we conservatively discard the
first $30\%$  of the point iterations at the beginning of any MCMC run, and thin the chains, we finally  extract the constraints on cosmological parameters by co-adding the thinned chains.
Histograms of the parameters from the merged chains were used to estimate the median values and confidence ranges.  In Tables (\ref{tab1}), and (\ref{tab2}) we report results of our analysis. It is worth noting that $\sigma_{int}$ is the observed dispersion of the QSO Hubble diagram, and it is of order of 0.2 dex. If we compare this value with dispersion in the Hubble diagram of SNIa, which is $\sigma_{SN}\sim 0.07$ at $z \sim 1$, it turns out that  QSOs provide the same cosmological information as SNIa. For the GRBs Hubble diagram we do not provide the observed dispersion here, because it has been evaluated during the calibration of the $E_{\rm p,i}$ -- $E_{\rm iso}$  correlation and used to estimates the error bars in the GRB Hubble diagram. In  Figs. (\ref{fig1}) , (\ref{fig2}) , and (\ref{fig3}) we show
$2D$ confidence regions:  it turns out that the dark energy EOS does evolve with redshift, and  the $\Lambda$CDM model, i.e. $w_0=-1$ and  $w_1=0$ is disfavored (at more
than $3 \sigma$) with all the
data, thus showing the importance of  independent and complementary data sets, specially in different ranges of redshift. Actually our results, not based on
cosmographic expansions, confirm the tension between predictions of the $\Lambda$CDM model and observations found previously in \cite{Lusso16, Risaliti19, Lusso19}.
In order to highlight that this result is due to the contribution of the high redshift QSO and GRB Hubble diagram we first used these sample only: the results are shown  in Table (\ref{tab2}) (right panel). In Figs. (\ref{residual}) and  (\ref{residual2}) we plot the distribution of the residuals as a function of redshift for the GRB and QSO samples. It tuns out that the amplitude of the scatter of the residuals show no significant  trend with redshift. Moreover  the deviation from the $\Lambda$CDM model  emerges at high redshifts: it turns out that if we select GRBs and QSOs at $z\leq 1$,  the corresponding Hubble diagram is well reproduced by the standard flat $\Lambda$CDM, as shown in Table (\ref{tablz}).\footnote{It is worth noting that the values of the QSOs and GRBs distance module are not absolute, thus  cross calibration parameters are needed (for both GRBs and QSOs data). Therefore the Hubble constant $H_0$ is degenerate with these calibration parameters, and it cannot be constrained.} Similarly when we fit the GRB and QSO Hubble diagram fixing $\Omega_m = 0.31$, as we show in Table (\ref{tabomfix}), we found that the $\Lambda$CDM with $\Omega_m= 0.31$  reproduces the data at $1 \sigma$.
 Finally it is worth noting that future missions, like, the THESEUS observatory
\cite{Amati_theseus} for GRBs, and eRosita all sky survey for QSOs (\cite{Lusso 20} ) will
 substantially increase the number of data usable to construct the Hubble diagram at high redshift and so help to probe the nature of dark energy. Actually the main  power of the GRB and QSO  Hubble diagram for cosmological investigations lies in the high-redshift regime, where it is possible  to discriminate among different cosmological models and also different theories of gravity.
  \begin{table*}
\begin{center}
%\scriptsize
\begin{tabular}{ccccccc}
%\hline
\, & \multicolumn{5}{c}{\bf CPL Dark Energy}   \\
\, & \, & \, & \, & \, & \,  \\
\hline
\, & \, & \, & \, & \, & \,   \\
$Id$ & $\langle x \rangle$ & $\tilde{x}$ & $68\% \ {\rm CL}$  & $95\% \ {\rm CL}$  \\
\hline \hline
\, & \, & \, & \, & \, & \, \\
\hline \, & \multicolumn{5}{c}{SNIa/GRBs/{\bf QSOs}/H(z)/BAO}
 \\
\hline
\, & \, & \, & \, & \, & \,   \\
$\Omega_m$ &0.29 &0.28& (0.26, 0.31) & (0.24, 0.33)   \\
\, & \, & \, & \, & \, & \,   \\
$w_0$ &-0.92& -0.92& (-1.1, -0.73) & (-1.24,  -0.67)   \\
\, & \, & \, & \, & \, & \, \\
$w_1$ &{\bf -0.77}&{\bf -0.71}& (-0.9,-0.4) & (-1, -0.3)  \\
\, & \, & \, & \, & \, & \,   \\
$h$ &0.73& 0.73 & (0.69, 0.74) & (0.65, 0.75)   \\
\, & \, & \, & \, & \, & \,   \\
$\mu_0$ &25.5& 25.5 & (25.44, 25.6) & (25.35, 25.7)    \\
\, & \, & \, & \, & \, & \,   \\
$\sigma_{int}$&{\bf 0.165}& {\bf 0.17} & {\bf (0.16 , 0.175)} & {\bf (0.15, 0.18) }\\
\, & \, & \, & \, & \, & \,  \\
\hline
\end{tabular}
\end{center}
\caption{Constraints on the CPL parameters from different data: combined SNIa, GRBs, and QSOs Hubble diagrams,
$H(z)$ data sets, and BAO data. Columns show the mean $\langle x \rangle$ and median $\tilde{x}$ values  and the $68\%$ and $95\%$
confidence limits.}
\label{tab1}
\end{table*}

 \begin{table*}
\begin{center}
%\scriptsize
\begin{tabular}{cccccccccc}
%\hline
\, & \multicolumn{7}{c}{\bf CPL Dark Energy}   \\
\, & \, & \, & \, & \, & \, & \, & \, & \,   \\
\hline
\, & \, & \, & \, & \, & \, & \, & \, & \,   \\
$Id$ & $\langle x \rangle$ & $\tilde{x}$ & $68\% \ {\rm CL}$  & $95\% \ {\rm CL}$ &  $\langle x \rangle$ & $\tilde{x}$ & $68\% \ {\rm CL}$  & $95\% \ {\rm CL}$ \\
\hline \hline
\, & \, & \, & \, & \, & \, & \, & \, & \,   \\
\hline \, & \multicolumn{4}{c}{SNIa/GRBs/H(z)}  \, & \multicolumn{4}{c}{QSOs/GRBs}
 \\
\hline
\, & \, & \, & \, & \, & \, & \, & \, & \,   \\
$\Omega_m$ &0.31 &0.31& (0.29, 0.33) & (0.25, 0.35) &0. 28&0.28& (0.26, 0.31) & (0.24, 0.33)  \\
\, & \, & \, & \, & \, & \, & \, & \, & \,  \\
$w_0$ &{\bf -1.05}& {\bf -1.07}& (-1.15, -0.80) & (-1.19,  -0.69)  &-1.1& -1.1& (-1.24, -0.93) & (-1.45, -0.81) \\
\, & \, & \, & \, & \, & \, & \, & \, & \, &  \\
$w_1$ &{\bf -0.64}&{\bf -0.65 }& (-0.90,-0.3) & (-0.98, -0.18) &-0.69& -0.7& (-0.83, -0.57) & (-0.97, -0.38)  \\
\, & \, & \, & \, & \, & \, & \, & \, & \, &  \\
$h$ &0.70& 0.70 & (0.68, 0.72) & (0.65, 0.74)  &--& -- & -- & -- \\
\, & \, & \, & \, & \, & \, & \, & \, & \, &  \\
$\mu_0$ &25.4& 25.4 & (25.39, 25.49) & (25.36, 25.5)  &25.5& 25.5 & (25.47, 25.54) & (25.42, 25.57)  \\
\, & \, & \, & \, & \, & \, & \, & \, & \, &  \\
${\mathbf \sigma}_{int}(dex)$ &--& -- & --& --&0.17& 0.17 &(0.16, 0.175) & (0.15, 0.18)  \\
\, & \, & \, & \, & \, & \, & \, & \, & \, &  \\
\hline
\end{tabular}
\end{center}
\caption{Constraints on the CPL parameters from different data: combined SNIa  and  GRB Hubble diagrams,  and
$H(z)$ data sets (Left Panel);  QSO and GRB Hubble diagram (Right Panel ). Columns show the mean $\langle x \rangle$ and median $\tilde{x}$ values  and the $68\%$ and $95\%$
confidence limits.}  \label{tab2}
\end{table*}

 \begin{table*}
\begin{center}
%\scriptsize
\begin{tabular}{ccccccc}
%\hline
\, & \multicolumn{5}{c}{\bf CPL Dark Energy}   \\
\, & \, & \, & \, & \, & \,  \\
\hline
\, & \, & \, & \, & \, & \,   \\
$Id$ & $\langle x \rangle$ & $\tilde{x}$ & $68\% \ {\rm CL}$  & $95\% \ {\rm CL}$  \\
\hline \hline
\, & \, & \, & \, & \, & \, \\
\hline \, & \multicolumn{5}{c}{GRBs/QSO-LZ}
 \\
\hline
\, & \, & \, & \, & \, & \,   \\
$\Omega_m$ &0.3 &0.31& (0.16, 0.41) & (0.12, 0.46)   \\
\, & \, & \, & \, & \, & \,   \\
$w_0$ &-1.03& -0.93& (-1.5, -0.6) & (-1.9,  -0.51)   \\
\, & \, & \, & \, & \, & \, \\
$w_1$ &{\bf -0.34}& -0.35& (-0.75,0.1) & (-1.4, 0.64)  \\
\, & \, & \, & \, & \, & \,   \\
$\mu_0$ &25.3& 25.3 & (25.1, 25.5) & (25. 25.7)    \\
\, & \, & \, & \, & \, & \,   \\
$\sigma_{int}$&0.175& {\bf 0.174} & {\bf (0.16 , 0.19)} & {\bf (0.15, 0.20) }\\
\, & \, & \, & \, & \, & \,  \\
\hline
\end{tabular}
\end{center}
\caption{Constraints on the CPL parameters from GRB and QSO Hubble diagram at $z\leq1$. Columns show the mean $\langle x \rangle$ and median $\tilde{x}$ values  and the $68\%$ and $95\%$
confidence limits.}
\label{tablz}
\end{table*}

 \begin{table*}
\begin{center}
%\scriptsize
\begin{tabular}{ccccccc}
%\hline
\, & \multicolumn{5}{c}{\bf CPL Dark Energy}   \\
\, & \, & \, & \, & \, & \,  \\
\hline
\, & \, & \, & \, & \, & \,   \\
$Id$ & $\langle x \rangle$ & $\tilde{x}$ & $68\% \ {\rm CL}$  & $95\% \ {\rm CL}$  \\
\hline \hline
\, & \, & \, & \, & \, & \, \\
\hline \, & \multicolumn{5}{c}{GRBs/QSO $\Omega_m$fixed}
 \\
\hline
\, & \, & \, & \, & \, & \,   \\
$w_0$ &-0.9& -0.93& (-1.15, -0.71) & (-1.4,  -0.61)   \\
\, & \, & \, & \, & \, & \, \\
$w_1$ &-0.03 & -0.05& (-0.25,0.22) & (-0.55, 0.46)  \\
\, & \, & \, & \, & \, & \,   \\
$\mu_0$ &25.6& 25.6 & (25.5, 25.7) & (25. 45,25.8)    \\
\, & \, & \, & \, & \, & \,   \\
$\sigma_{int}$&0.17& {\bf 0.175} & {\bf (0.16 , 0.2)} & {\bf (0.15, 0.24) }\\
\, & \, & \, & \, & \, & \,  \\
\hline
\end{tabular}
\end{center}
\caption{Constraints from GRB and QSO Hubble diagram, setting $\Omega_m=0.31$ . Columns show the mean $\langle x \rangle$ and median $\tilde{x}$ values  and the $68\%$ and $95\%$
confidence limits. }
\label{tabomfix}
\end{table*}

\section{Conclusions}
The nature of the dark energy and the origin of the accelerated expansion of the Universe remain one of the most challenging open questions in Physics and Cosmology.
The flat $\Lambda$CDM model is the most popular cosmological model  used by the scientific community. However, despite its enormous successes, some problems have been detected: quite recently it turned out that there is a tension (at more than $3 \sigma$)  between cosmological and local  measurements of the Hubble constant (\citep{Riess19} ).
Moreover, recently in \citep{Lusso19,Risaliti19,Lusso16,MEDL19}, another tension has been also reported
between predictions of the $\Lambda$CDM model and the Hubble Diagram of SNIa and quasars. With the aim of
clarifying this result, we concentrated on the possibility to detect, from different and independent data, evidence of a redshift evolution of the dark energy equation of state: we performed  statistical
analysis to constrain the dark energy EOS, using the simple CPL parametrization. Our high redshift Hubble diagram provides a clear indication (at $3\sigma$ level) of an evolving dark energy EOS, thus confirming the previous results \cite{MEDL19}, and highlight the importance, to explore the cosmological expansion, of using independent probes and exploring large ranges of redshift. It turns out that the deviation from the standard $\Lambda$CDM is due just to the QSO and GRBs Hubble diagrams. Moreover it is important for $ z > 1$: if, indeed we limit our analysis to $z\leq1$ the $\Lambda$CDM  reproduces the data at $1 \sigma$.   The residuals do not present any significant trend systematic with redshift: this evidence further proves that our results are not affected by systematics.

With  future missions, like, the THESEUS observatory, and the eRosita all-sky survey, that will substantially increase the number of GRBs and QSOs usable to construct the Hubble diagram it will be possible to  better test  the nature of dark energy.
%\begin{figure}[h!] 
%\begin{center}
%\includegraphics[width=8cm]{w0w12dsim.eps}% This is a *.eps file
%\end{center}
%\caption{ 2D confidence regions in the $w_0-w_1$ plane for the CPL model, obtained from 772 simulated GRBs, QSOs and the $H(z)$ measurements. }\label{fig3}
%\end{figure}

%\section*{Author Contributions}
%EP has written most of the text. All the authors contributed to the analysis and critically reviewed the entire paper.
\section*{Funding}
We acknowledge financial contribution from the agreement ASI-INAF n.2017-14-H.O.
MD is grateful to the INFN for financial support through the Fondi FAI GrIV.

\section*{Acknowledgments}
EP acknowledges the support of INFN Sez. di Napoli  (Iniziativa Specifica QGSKY ).

\bibliographystyle{plain}

\end{document}